\documentclass{PoS}

\title{On the smallness of the cosmological constant in SUGRA models with Planck scale SUSY breaking and degenerate vacua}

\ShortTitle{On the smallness of the cosmological constant
}

\author{Colin Froggatt\\
       Glasgow University\\
       E-mail: \email{Colin.Froggatt@glasgow.ac.uk}}

\author{\speaker{Roman Nevzorov}%
         \thanks{On leave of absence from the Theory Department, ITEP, Moscow, Russia.}\\
        University of Adelaide \\
        E-mail: \email{roman.nevzorov@adelaide.edu.au}}

\author{Holger Bech Nielsen\\
        The Niels Bohr Institute\\
        E-mail: \email{hbech@nbi.dk}}

\author{Anthony Thomas\\
        University of Adelaide\\
        E-mail: \email{anthony.thomas@adelaide.edu.au}}

\abstract{
We argue that the exact degeneracy of vacua in $N=1$ supergravity
can shed light on the smallness of the cosmological constant. The presence
of such vacua, which are degenerate to very high accuracy, may also result
in small values of the quartic Higgs coupling and its beta function at the
Planck scale in the phase in which we live.}

\FullConference{The European Physical Society Conference on High Energy Physics\\
                 22-29 July 2015\\
                 Vienna, Austria}

\begin{document}

\section{Higgs mass and the multiple point principle}

The recent discovery of the Higgs boson allows us to determine quite precisely the parameters
of the Higgs potential of the standard model (SM)
\begin{equation}
V_{eff}(H) = m^2(\phi) H^{\dagger} H + \lambda(\phi) (H^{\dagger} H)^2\,.
\label{1}
\end{equation}
Here $H$ is a Higgs doublet and $\phi$ is a norm of the Higgs field, i.e. $\phi^2=H^{\dagger} H$.
The measured mass of the SM--like Higgs state is around $M_H\simeq 125-126\,\mbox{GeV}$,
which corresponds to a rather small value of the Higgs quartic coupling at the electroweak (EW) scale,
i.e. $\lambda\simeq 0.13$. Of critical importance here is the observation that this value of $\lambda$
is very close to its theoretical lower bound, coming from the vacuum stability constraint.
Indeed, for sufficiently small values of the Higgs quartic coupling at the EW scale, $\lambda(\phi)$
decreases with increasing energy scale and can become negative at some intermediate scale,
resulting in either instability or metastability of the physical vacuum. The extrapolation of the SM couplings
up to the Planck scale, $M_{Pl}$, using 3--loop renormalization group equations (RGEs) leads to
(see, for example \cite{Buttazzo:2013uya})
\begin{eqnarray}
\lambda(M_{Pl}) =  -0.0143-0.0066\left(\frac{M_t}{\mbox{GeV}} - 173.34 \right) \qquad\qquad\qquad\qquad\qquad\qquad\qquad \nonumber\\[2mm]
\qquad\qquad\qquad\qquad+0.0018\left(\frac{\alpha_3(M_Z)-0.1184}{0.0007}\right) +0.0029\left(\frac{M_H}{\mbox{GeV}} -125.15 \right)\,,
\label{2}
\end{eqnarray}
where $M_t$ is the top quark mass and $\alpha_3(M_Z)$ is the value of the strong gauge coupling at
the EW scale. Eq.~(\ref{2}) indicates that the value of $\lambda(M_{Pl})$ tends to be rather close to zero
for any phenomenologically acceptable values of $M_t$ and $\alpha_3(M_Z)$. Moreover, $\lambda(M_{Pl})$
can be positive if the top quark is lighter than $171\,\mbox{GeV}$. This means that for these values of
the top quark pole mass $\lambda(\phi)$ remains positive at any intermediate scale below $M_{Pl}$,
so that the physical vacuum is stable and the parameters of the SM can be extrapolated all the way up to $M_{Pl}$
without any inconsistency.

For $M_t\simeq 171\,\mbox{GeV}$ the value of $\lambda(M_{Pl})$ vanishes. In this context it is worth noting
that the computed value of  the beta--function of $\lambda(\phi)$, $\beta_{\lambda}=\frac{d \lambda(\phi)}{d \log\phi}$, 
also tends to be very small near the Planck scale. This implies that the Higgs effective potential of the SM (\ref{1}) has two rings
of minima in the Mexican hat with the same vacuum energy density. The radius of the little ring equals the EW vacuum
expectation value (VEV) of the Higgs field, while the radius of the second ring is somewhat close to the Planck scale.

The presence of such degenerate vacua was predicted by the so-called Multiple Point Principle (MPP)
\cite{Bennett:1993pj}-\cite{Froggatt:1995rt}, according to which Nature chooses values of coupling constants
such that many phases of the underlying theory should coexist. This scenario corresponds to a special (multiple)
point on the phase diagram of the theory where these phases meet. The vacuum energy densities of these different
phases are degenerate at the multiple point. The MPP applied to the SM implies that the Higgs effective potential has
two degenerate vacua which are taken to be at the EW and Planck scales \cite{Froggatt:1995rt}.
The degeneracy of these vacua can be achieved only if
\begin{equation}
\lambda(M_{Pl})\simeq 0 \,, \qquad\quad \beta_{\lambda}(M_{Pl})\simeq 0\,.
\label{3}
\end{equation}
It was shown that the MPP conditions (\ref{3}) are fulfilled when $M_t=173\pm 5\,\mbox{GeV}$ and
$M_H=135\pm 9\, \mbox{GeV}$ \cite{Froggatt:1995rt}. In previous articles the application of the MPP to the two Higgs
doublet extension of the SM was studied \cite{Froggatt:2006zc}--\cite{Froggatt:2008am}. In particular, it was pointed out that the MPP
can be used as a mechanism for the suppression of the flavour changing neutral current and CP--violation effects \cite{Froggatt:2008am}.

The success of the MPP in predicting the Higgs mass suggests that one can also use it to provide an explanation for the
small deviation of the cosmological constant from zero. This can be achieved by adapting the MPP to models based
on $(N=1)$ local supersymmetry -- supergravity (SUGRA),

\section{No-scale inspired SUGRA model with degenerate vacua}

The Lagrangian of $(N=1)$ supergravity is specified in terms of a real gauge-invariant K$\ddot{a}$hler function
$G(\phi_{M},\phi_{M}^{*})$ and an analytic gauge kinetic function $f_a(\phi_{M})$, which depend on the chiral superfields
$\phi_M$. The function $f_{a}(\phi_M)$ determines the kinetic terms for the fields in the vector supermultiplets and the gauge
coupling constants $Re f_a(\phi_M)=1/g_a^2$, where the index $a$ designates different gauge groups. The K$\ddot{a}$hler function is
given by
\begin{equation}
G(\phi_{M},\phi_{M}^{*})=K(\phi_{M},\phi_{M}^{*})+\ln|W(\phi_M)|^2\,,
\label{4}
\end{equation}
where $K(\phi_{M},\phi_{M}^{*})$ and $W(\phi_M)$ are the K$\ddot{a}$hler potential and superpotential of the SUGRA model under consideration.
Here we set $\frac{M_{Pl}}{\sqrt{8\pi}}=1$. In $(N=1)$ SUGRA models the breakdown of local SUSY takes place in a hidden sector that
contains superfields $z_i$, which are singlets under the SM gauge group. The absolute value of the vacuum energy density
at the minimum of the SUGRA scalar potential tends to be of order of $m_{3/2}^2 M_{Pl}^2$, where $m_{3/2}$ is a gravitino mass.
Thus an enormous fine--tuning is required to keep the vacuum energy density in SUGRA models around the observed value of the cosmological
constant $\rho_{\Lambda} \sim 10^{-123}M_{Pl}^4$ \cite{Bennett:2003bz}.

The situation changes substantially in no-scale supergravity, where the invariance of the Lagrangian under
imaginary translations and dilatations leads to the vanishing of the vacuum energy density. However
these global symmetries also protect supersymmetry which must be broken. The breakdown of dilatation
invariance does not necessarily result in a non--zero vacuum energy density \cite{Froggatt:2005nb}.
Let us consider the case when the dilatation invariance is broken
in the superpotential of the hidden sector only. The simplest hidden sector
of a SUGRA model of this type includes two superfields, $T$ and $z$, which transform differently under
the imaginary translations ($T\to T+i\beta,\, z\to z$) and dilatations ($T\to\alpha^2 T,\, z\to\alpha\,z$).
If the superpotential and K$\ddot{a}$hler potential of the hidden sector of this SUGRA model are given by
\begin{equation}
K(T,\,z)=-3\ln\biggl[T+\overline{T}-|z|^2\biggr]\,,\qquad
W(z)=\kappa\biggl(z^3+ \mu_0 z^2 \biggr)\,,
\label{8}
\end{equation}
then the tree level scalar potential of the hidden sector is positive definite
\begin{equation}
V(T,\, z)=\frac{1}{3(T+\overline{T}-|z|^2)^2}
\biggl|\frac{\partial W(z)}{\partial z}\biggr|^2\,,
\label{9}
\end{equation}
so that the vacuum energy density vanishes near its global minima. The bilinear mass term for
the superfield $z$ in the superpotential $W(z)$ violates dilatation invariance. The SUGRA scalar
potential (\ref{9}) possesses two minima  at $z=0$ and $z=-\frac{2\mu_0}{3}$ that correspond
to the stationary points of the hidden sector superpotential. In the first vacuum, where $z=-\frac{2\mu_0}{3}$,
local SUSY is broken and the gravitino becomes massive. Because, in general, $\mu_0\lesssim M_{Pl}$ and
$\kappa\lesssim 1$ SUSY is broken in this vacuum near the Planck scale. In the second minimum,
with $z=0$, the gravitino mass vanishes and local SUSY remains intact.

The SUGRA model discussed above leads to a natural realisation of the MPP.
The successful implementation of the MPP in $(N=1)$ supergravity requires the existence
of a vacuum in which the low--energy limit of this theory is described by a pure supersymmetric model
in flat Minkowski space. According to the MPP this vacuum and the physical one must be
degenerate \cite{Froggatt:2005nb}--\cite{Froggatt:2014jza}. Such a second vacuum is
only realised if the SUGRA scalar potential has a minimum where $m_{3/2}=0$.
This would normally requires an extra fine-tuning \cite{Froggatt:2003jm}. In the SUGRA model with
the superpotential and K$\ddot{a}$hler potential (\ref{8}) the MPP conditions are fulfilled automatically,
without any extra fine-tuning at the tree--level. Quantum corrections to the
Lagrangian of this no--scale inspired SUGRA model are expected to spoil the
degeneracy of the vacua, giving rise to a huge energy density in the vacuum where SUSY
is broken. Therefore such model should be considered as a toy example
only. It demonstrates that, in $(N=1)$ supergravity, there might be a mechanism which
ensures the vanishing of vacuum energy density in the physical vacuum. Such a mechanism
can also result in a set of degenerate supersymmetric and non-supersymmetric Minkowski
vacua, leading to the realization of the multiple point principle.

\section{Implications for cosmology and Higgs phenomenology}

Let us now assume that a SUGRA model with at least two vacua, which are exactly degenerate,
is realised in Nature. In the first (physical) vacuum the spontaneous breakdown of SUSY takes
place near the Planck scale. In the second vacuum the low--energy limit of the theory is
described by a pure SUSY model in flat Minkowski space. Because the vacuum
energy density of supersymmetric states in flat Minkowski space vanishes
and all vacua in the MPP inspired SUGRA models are degenerate, the cosmological constant
problem in the physical vacuum is solved to first approximation by our assumption. At the same
time non--perturbative effects in the hidden sector can lead to the breakdown of SUSY in the
supersymmetric phase at low energies, resulting in a small vacuum energy density.
This small value should be then transferred to our vacuum by the assumed degeneracy.

Further, we assume that in the SUSY Minkowski vacuum vector supermultiplets associated with the
unbroken gauge symmetry in the hidden sector remain massless. These vector supermultiplets
may give rise to the breakdown of SUSY near the scale $\Lambda_{SQCD}$, where the supersymmetric
QCD type interactions in the hidden sector become strong in this second vacuum. The appropriate
breakdown of supersymmetry can be caused by the formation of a gaugino condensate. This condensate
itself does not break global SUSY. However the inclusion of non-renormalisable terms in the
Lagrangian of the $(N=1)$ SUGRA model can result in a non-trivial dependence of the gauge
kinetic function $f_{X}(z_m)$ on the hidden sector superfields $z_m$. As a consequence, auxiliary fields
$F^{z_m}$ associated with superfields $z_m$ acquire non--zero VEVs, which are set by
$<\bar{\lambda}_a\lambda_a>\simeq \Lambda_{SQCD}^3$, i.e.
\begin{equation}
F^{z_m}\propto \frac{\partial f_X(z_k)}{\partial z_m}\bar{\lambda}_a\lambda_a \sim \frac{\Lambda^3_{SQCD}}{M_{Pl}}\,.
\label{16}
\end{equation}
Since the breakdown of local SUSY is caused by non-renormalisable terms that are suppressed
by an inverse power of the Planck scale $M_{Pl}$, the gaugino condensate gives rise to a vacuum energy density
\begin{equation}
\rho^{(2)}_{\Lambda} \sim \frac{\Lambda_{SQCD}^6}{M_{Pl}^2}\, ,
\label{17}
\end{equation}
which is many orders of magnitude lower than $\Lambda_{SQCD}^4$.

Because of the postulated exact degeneracy of vacua, the physical phase, where we live, should have the same
energy density as the phase in which SUSY breaking is induced by the gaugino condensate in the hidden sector.
Then from Eq.~(\ref{17}) it follows that one can reproduce the observed value of the dark energy density if
$\Lambda_{SQCD}$ is relatively close to $\Lambda_{QCD}$ in the physical vacuum \cite{Froggatt:2014jza}, i.e.
\begin{equation}
\Lambda_{SQCD}\sim \Lambda_{QCD}/10\,.
\label{18}
\end{equation}
Although there is no compelling reason to expect that these two scales should be relatively close,
$\Lambda_{QCD}$ and $M_{Pl}$ can be considered as the two most natural choices
for the scale of dimensional transmutation in the hidden sector in the second vacuum.
In the case when the QCD type interactions in the hidden sector are described by $SU(3)$ SUSY gluodynamics
the corresponding value of $\Lambda_{SQCD}$ can be obtained when the $SU(3)$ gauge coupling
$g_X(M_{Pl})\simeq 0.654$, which is just slightly larger than the value of the QCD gauge coupling at the Planck
scale in the SM, i.e. $g_3(M_{Pl})=0.487$.

In principle, there may also be a third phase, which has the same energy density
as the first and second phases. In this third vacuum local SUSY and EW symmetry
can be broken somewhere near $M_{Pl}$. Since now the Higgs VEV is rather close
to $M_{Pl}$ one has to take into account the interaction of the Higgs and hidden
sector fields. Nonetheless, such interactions can be very weak if the VEV of the Higgs
field is considerably smaller than $M_{Pl}$ (say $\langle H \rangle \sim M_{Pl}/10$)
and the couplings of the SM Higgs doublet to the hidden sector fields are suppressed.
Then the VEVs of the hidden sector fields in the third and physical vacua can be
almost identical. As a consequence, one can expect that the gauge couplings and
$\lambda(M_{Pl})$ in the first and third phases are basically the same and the
absolute value of $m^2$ in the Higgs effective potential is much smaller than
$M_{Pl}^2$ and $\langle H^{\dagger} H\rangle$ in the third vacuum. Therefore
the existence of such a third vacuum with vanishingly small energy density again
implies that $\lambda(M_{Pl})$ and $\beta_{\lambda}(M_{Pl})$ are approximately
zero in the third vacuum. Because the couplings in the first and third phases are
basically identical, the presence of such a third vacuum leads to
$\lambda(M_{Pl})\approx \beta_{\lambda}(M_{Pl})\approx 0$ in the physical vacuum.

The estimation of the dark energy density discussed here is based on the assumption that
the vacua under consideration are degenerate to extremely high accuracy. In fact, the required
accuracy should be of the order of the value of the cosmological constant in the physical vacuum.
The desired accuracy can be achieved if the underlying theory allows only vacua which lead to
a similar order of magnitude of space-time 4-volumes of the Universe at its final stage of
evolution. Because the sizes of these volumes are determined by the expansion rates of the
vacua associated with them, all allowed vacua should have energy densities of the same order
of magnitude, i.e. of order of the observed value of the dark energy density in the phase where we live.

\acknowledgments
This work was supported by the University of Adelaide and the Australian Research Council through the ARC
Center of Excellence in Particle Physics at the Terascale and through grant LF0 99 2247 (AWT). HBN thanks
the Niels Bohr Institute for his emeritus status. CDF thanks Glasgow University and the Niels Bohr Institute
for hospitality and support.


\begin{thebibliography}{99}

\bibitem{Buttazzo:2013uya}
D. Buttazzo, G. Degrassi, P. P. Giardino, G. F. Giudice, F. Sala, A. Salvio, A. Strumia,
\emph{Investigating the near-criticality of the Higgs boson},
\emph{JHEP} {\bf 1312} (2013) 089 [{\tt arXiv:1307.3536 [hep-ph]}].

\bibitem{Bennett:1993pj}
D. L. Bennett, H. B. Nielsen,
\emph{Predictions for nonAbelian fine structure constants from multicriticality},
\emph{ Int.\ J.\ Mod.\ Phys.\ A} {\bf 9} (1994) 5155 [{\tt hep-ph/9311321}].

\bibitem{Froggatt:1995rt}
C. D. Froggatt, H. B. Nielsen,
\emph{Standard model criticality prediction: Top mass 173 +- 5-GeV and Higgs mass 135 +- 9-GeV},
\emph{Phys.\ Lett.\ B} {\bf 368} (1996) 96 [{\tt hep-ph/9511371}].

\bibitem{Froggatt:2006zc}
C. D. Froggatt, L. Laperashvili, R. Nevzorov, H. B. Nielsen, M.~Sher,
\emph{Implementation of the multiple point principle in the two-Higgs doublet model of type II},
\emph{Phys.\ Rev.\ D} {\bf 73} (2006) 095005 [{\tt hep-ph/0602054}].

\bibitem{Froggatt:2007qp}
C. D. Froggatt, R. Nevzorov, H. B. Nielsen, D. Thompson,
\emph{Fixed point scenario in the Two Higgs Doublet Model inspired by degenerate vacua},
\emph{Phys.\ Lett.\ B} {\bf 657} (2007) 95 [{\tt arXiv:0708.2903 [hep-ph]}].

\bibitem{Froggatt:2008am}
C. D. Froggatt, R. Nevzorov, H. B. Nielsen, D. Thompson,
\emph{On the origin of approximate custodial symmetry in the Two-Higgs Doublet Model},
\emph{Int.\ J.\ Mod.\ Phys.\ A} {\bf 24} (2009) 5587 [{\tt arXiv:0806.3190 [hep-ph]}].

\bibitem{Bennett:2003bz}
C.L. Bennett {\it et al.} [WMAP Collaboration],
\emph{First year Wilkinson Microwave Anisotropy Probe (WMAP) observations: Preliminary maps and basic results},
\emph{Astrophys.\ J.\ Suppl.} {\bf 148} (2003) 1 [{\tt astro-ph/0302207}].


\bibitem{Froggatt:2005nb}
C. Froggatt, R. Nevzorov, H.B. Nielsen,
\emph{On the smallness of the cosmological constant in SUGRA models},
\emph{Nucl.\ Phys.\ B} {\bf 743} (2006) 133 [{\tt hep-ph/0511259}].

\bibitem{Froggatt:2003jm}
C. Froggatt, L. Laperashvili, R. Nevzorov, H.B. Nielsen,
\emph{Cosmological constant in SUGRA models and the multiple point principle},
\emph{Phys.\ Atom.\ Nucl.} {\bf 67} (2004) 582 [{\tt hep-ph/0310127}].

\bibitem{Froggatt:2007qs}
C.D. Froggatt, R. Nevzorov, H.B. Nielsen,
\emph{Smallness of the cosmological constant and the multiple point principle},
in proceedings of \emph{2007 Europhysics Conference on High Energy Physics (EPS-HEP 2007)},
\emph{ J.\ Phys.\ Conf.\ Ser.}  {\bf 110} (2008) 072012 [{\tt arXiv:0708.2907 [hep-ph]}].


\bibitem{Froggatt:2009pj}
C.D. Froggatt, R. Nevzorov, H.B. Nielsen,
\emph{On the Smallness of the Cosmological Constant in SUGRA Models Inspired by Degenerate Vacua},
in proceedings of \emph{17th International Conference on Supersymmetry and the Unification of Fundamental Interactions (SUSY 2009)}
\emph{AIP Conf.\ Proc.}  {\bf 1200} (2010) 1093 [{\tt arXiv:0909.4703 [hep-ph]}].

\bibitem{Froggatt:2010iy}
C. Froggatt, R. Nevzorov, H.B. Nielsen,
\emph{Dark Energy density in Split SUSY models inspired by degenerate vacua},
in proceedings of \emph{35th International Conference on High Energy Physics (ICHEP 2010)},
\pos{PoS(ICHEP2010)442} (2010) [{\tt arXiv:1012.5121 [hep-ph]}].

\bibitem{Froggatt:2011fc}
C. Froggatt, R. Nevzorov, H.B. Nielsen,
\emph{Dark Energy density in models with Split Supersymmetry and degenerate vacua},
\emph{Int.\ J.\ Mod.\ Phys.\ A} {\bf 27} (2012) 1250063 [{\tt arXiv:1103.2146 [hep-ph]}].


\bibitem{Froggatt:2014jza}
C.D. Froggatt, R. Nevzorov, H.B. Nielsen, A.W. Thomas,
\emph{Cosmological constant in SUGRA models with Planck scale SUSY breaking and degenerate vacua},
\emph{Phys.\ Lett.\ B} {\bf 737} (2014) 167 [{\tt arXiv:1403.1001 [hep-ph]}].



\end{thebibliography}
\end{document}